\documentclass[twocolumn,showpacs,aps]{revtex4}
\usepackage{epsf}
\begin{document}
\title{Size of Outbreaks Near the Epidemic Threshold}
\author{E.~Ben-Naim}\email{ebn@lanl.gov} \affiliation{Theoretical
Division and Center for Nonlinear Studies, Los Alamos National
Laboratory, Los Alamos, NM, 87545}
\author{P.~L.~Krapivsky}\email{paulk@bu.edu} \affiliation{Center for
BioDynamics and Department of Physics, Boston University, Boston, MA,
02215}
\begin{abstract}
  
  The spread of infectious diseases near the epidemic threshold is
  investigated.  Scaling laws for the size and the duration of
  outbreaks originating from a single infected individual in a large
  susceptible population are obtained.  The maximal size of an
  outbreak $n_*$ scales as $N^{2/3}$ with $N$ the population size.
  This scaling law implies that the average outbreak size $\langle
  n\rangle$ scales as $N^{1/3}$.  Moreover, the maximal and the
  average duration of an outbreak grow as $t_*\sim N^{1/3}$ and
  $\langle t\rangle\sim \ln N$, respectively.

\end{abstract}


\maketitle

Infection processes typically involve a threshold
\cite{ntjb,ma,hwh,wd,hed,lp,bc,ad,jg}.  Below the epidemic threshold,
outbreaks quickly die out, while above the threshold, outbreaks may
take off.  We study epidemic outbreaks near the threshold. Such
outbreaks arise naturally.  On the one hand, human efforts at disease
prevention reduce the infection rate thereby crossing the epidemic
threshold \cite{ma}. On the other hand, evolution may increase the
infection rate of diseases hovering just below the threshold,
enhancing the likelihood of near-threshold outbreaks
\cite{antia}. Typically, detection, modeling, and eradication of
infectious diseases are subtle for outbreaks near the epidemic
threshold.

The total number of infected individuals is a basic measure of the
severity of an epidemic outbreak. We study outbreaks originating from 
a single infected individual in a large susceptible population.  Our
main result is that near the epidemic threshold, the maximal outbreak
size $n_*$ grows as a power-law of the population size $N$,
\begin{eqnarray}
\label{size-max}
n_*\sim N^{2/3}\,.
\end{eqnarray}
In contrast, below the epidemic threshold, endemic outbreaks involve a
small number of infected individuals, while above the epidemic
threshold, pandemic outbreaks involve a fraction of the population
$n_*\sim N$. Therefore, outbreaks near the epidemic threshold have a
distinct intermediate size between a pandemic and an endemic outbreak
\cite{ntjb}.  Loosely speaking, epidemics come in three sizes: large,
medium, and small.

The scaling law (\ref{size-max}) has several important implications
concerning the statistics of both the size and the duration of the
outbreaks.  It implies that the average size of outbreaks $\langle
n\rangle$ and the maximal duration of outbreaks $t_*$ both scale as
$\langle n\rangle\sim t_*\sim N^{1/3}$ near the epidemic
threshold. Furthermore, the average duration of the outbreaks $\langle
t\rangle$ scales logarithmically, $\langle t\rangle\sim \ln N$. These
behaviors hold in a sizable range of infection rates, namely in a
window of the order ${\cal O}(N^{-1/3})$ around the epidemic
threshold.

These scaling laws are demonstrated for the classic
Susceptible-Infected-Recovered (SIR) infection process
\cite{ntjb,ma,hwh}. In this model, the population consists of $s$
susceptible, $i$ infected, and $r$ recovered individuals with
$N=s+i+r$.  These sub-populations change due to two competing
processes: infection and recovery. The disease is transmitted from an
infected individual to a susceptible one with rate $\alpha/N$, where
$\alpha$ is the infection rate:
\begin{equation}
(s,i,r)\buildrel {\alpha si}/N \over \longrightarrow (s-1,i+1,r).
\end{equation}
Infected individuals recover with a unit rate:
\begin{equation}
(s,i,r)\buildrel i \over \longrightarrow (s,i-1,r+1).
\end{equation}
The infection process starts with a single infected individual,
$(s,i,r)=(N-1,1,0)$, and it ends with none $(s,i,r)=(N-n,0,n)$.

\begin{figure}[t]
 \centerline{\epsfxsize=8cm\epsfbox{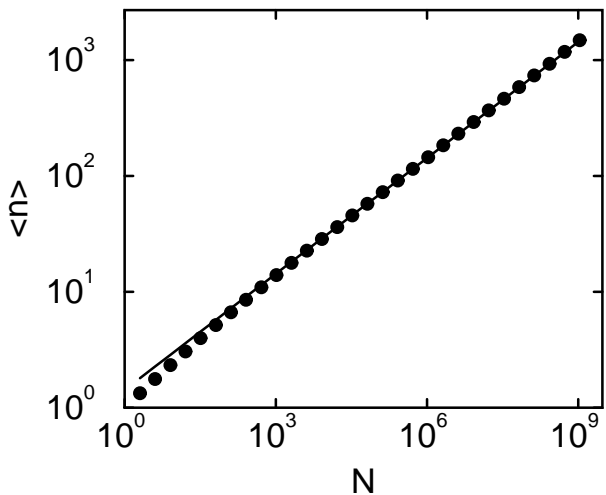}}
 \caption{The average outbreak size versus the population size for the
 SIR infection process at the epidemic threshold ($\alpha=1$). Shown
 are Monte Carlo simulation results representing an average over
 $10^9$ independent realizations of the infection process (circles). A
 line of slope $1/3$ is also shown as a reference. A least-square-fit
 to $\langle n\rangle \sim N^\gamma$ in the range $10^3<N<10^9$ yields
 $\gamma=0.334\pm 0.001$.}
\end{figure}

The total size of the outbreak $n$ and the duration of the outbreak
$t$ are the outcomes of a stochastic process. We study statistical
properties of these random variables, particularly, their average and
maximal size, as a function of the population size (We implicitly
consider an average over infinitely many realizations of the infection
process.)

In the infinite population limit, the epidemic threshold is
$\alpha=1$. Since infection occurs with probability
$\frac{\alpha}{1+\alpha}$ and recovery with probability
$\frac{1}{1+\alpha}$, the average outbreak size satisfies
\hbox{$\langle n\rangle
=\frac{1}{1+\alpha}+2\frac{\alpha}{1+\alpha}\langle n\rangle$}.  Thus,
below the threshold ($\alpha<1$), a finite number of individuals is
infected, $\langle n\rangle=(1-\alpha)^{-1}$. Above the threshold
($\alpha>1$), there is a pandemic outbreak with a finite fraction of
the population infected: $\langle n\rangle =rN$ \cite{ntjb,ar}. At the
threshold ($\alpha=1$), the probability that the outbreak size equals
$n$, $G_n$, is found recursively:
\hbox{$G_n=\frac{1}{2}\sum_{m=1}^{n-1} G_mG_{n-m}$} starting with
$G_1=1/2$. This recursion reflects the fact that the first infection
event results in two independent infection processes \cite{teh}.  The
generating function is \hbox{$\sum_{n\geq 1} G_nz^n =1-\sqrt{1-z}$},
from which the size distribution is a power-law,
\begin{equation}
\label{size-dist}
G_n\sim n^{-3/2},
\end{equation}
for sufficiently large outbreaks $n\gg 1$.

For a finite, yet large population, the outbreak size distribution
(\ref{size-dist}) holds, but only up to the maximal outbreak size:
$1\ll n \ll n_*$. Outbreaks beyond the maximal size are practically
impossible.  Therefore, the average outbreak size grows according to
$\langle n\rangle =\sum_{n=1}^{n_*} n G_n\sim n^{1/2}_*$. Naively
assuming that a finite fraction of the population may become infected,
$n_*\sim N$, would lead to \hbox{$\langle n\rangle \sim
N^{1/2}$}. While consistent with the generic statistical
uncertainties, this law is in fact {\it erroneous}. Instead, the
outbreak size is much smaller because the epidemic outbreak weakens as
more individuals become infected, and it finally dies out when the
number of infected individuals becomes of the order $n_*$. When there
are \hbox{$N-n_*$} susceptible individuals, the total infection rate
$\alpha (N-n_*)i/N$ shows that the infection rate is effectively
reduced, \hbox{$\alpha_{\rm eff}=1-n_*/N$}. Therefore, the epidemic
becomes essentially endemic. (This is clearly a finite population
effect: the susceptible population ``reservoir'' is never affected in
the infinite population limit.)  Equating the outbreak size in the
endemic phase \hbox{$\langle n\rangle\sim (1-\alpha_{\rm
eff})^{-1}\sim N/n_*$} with that estimated from the size distribution,
$\langle n\rangle\sim n_*^{1/2}$, gives the scaling law
(\ref{size-max}) governing the maximal outbreak size. Hence, in the
worse case scenario, only a fraction of the order of $N^{-1/3}$ of the
entire population can ever be infected.

As a byproduct we obtain the scaling law for the average outbreak size
\begin{eqnarray}
\label{size-ave}
\langle n\rangle\sim N^{1/3}\,.
\end{eqnarray}
Large scale Monte Carlo simulations confirm this behavior (Fig.~1).
The simulations are a straightforward realization of the infection
process. When there are $s$ susceptible individuals, with probability
$1/(1+\alpha s/N)$ a recovery event occurs, and otherwise, an
infection event occurs. The simulation results represent an average
over a remarkably large number of independent realizations.

\begin{figure}[t]
 \centerline{\epsfxsize=8cm\epsfbox{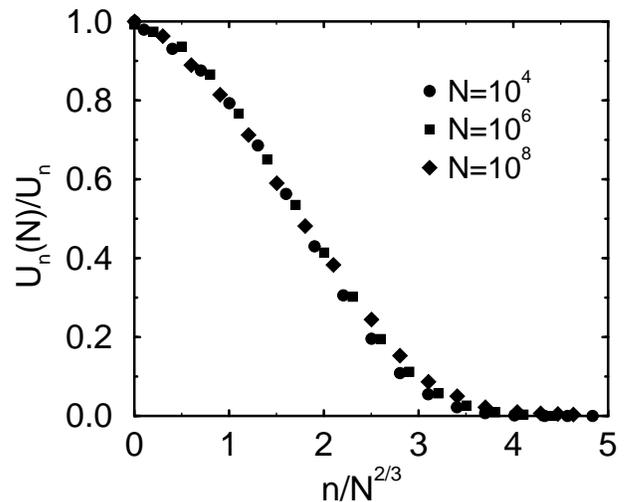}} \caption{The normalized
 cumulative distribution $U_n(N)/U_n(\infty)$ versus the normalized
 outbreak size $n/N^{2/3}$. The data corresponds to an average over
 $10^6$ independent realizations.}
\end{figure}

Statistical properties of the outbreak size are self-similar as they
follow a universal, population-size independent law.  Once the
outbreak size distribution and the outbreak size are properly
normalized by the infinite population distribution and the maximal
outbreak size respectively, a universal behavior emerges:
\hbox{$G_n(N)/G_n(\infty)\to {\cal G}(n/N^{2/3})$}. This universality,
reminiscent of finite-size scaling in critical phenomena \cite{pf},
was confirmed numerically by studying the cumulative distribution
$U_n(N)=\sum_{m\geq n}G_n(N)$ (Fig.~2). This provides further
verification of the scaling law (\ref{size-max}).

The scaling laws characterizing the outbreak size hold not only at the
threshold but also in a window around the threshold. Equating the
average outbreak size (\ref{size-ave}) with the behavior in the
endemic phase, $\langle n\rangle=(1-\alpha)^{-1}$, we find that the
threshold window (i.e., the range of infection rates for which the
intermediate behavior holds) diminishes with the population size as
\begin{equation}
\label{window}
|1-\alpha|\sim N^{-1/3}.
\end{equation}
This parameter range can be sizable for moderate populations --- for
example, when $N=10^3$, the threshold window is roughly
$0.9<\alpha<1.1$ and the maximal outbreak size is smaller than the
population size by a factor of $10$.

\begin{figure}[t]
  \centerline{\epsfxsize=8cm\epsfbox{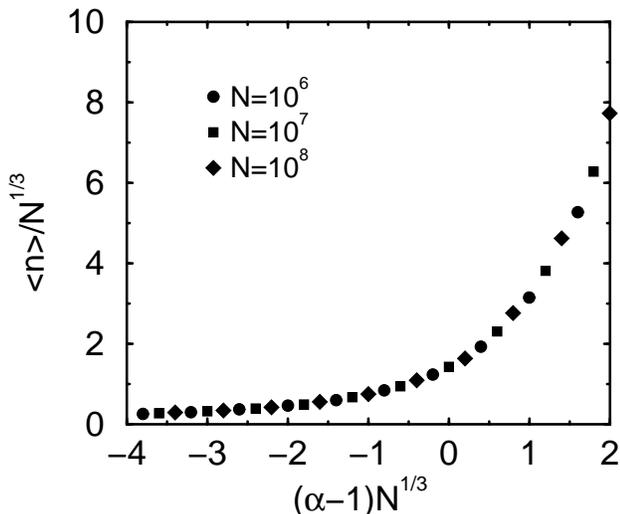}}
 \caption{The near threshold behavior. Shown is the normalized outbreak
 size $\langle n\rangle/N^{1/3}$ versus the normalized distance from
 the threshold $(\alpha-1)N^{1/3}$. The data corresponds to an average
 over $10^6$ independent realizations.}
\end{figure}

The behavior of $\langle n\rangle$ near the epidemic threshold
provides another manifestation of the scaling law
(\ref{window}). Indeed, plotting the average outbreak size versus the
infection rate normalized according to (\ref{size-ave}) and
(\ref{window}), respectively, shows a universal behavior:
\hbox{$\langle n\rangle/N^{1/3}\to {\cal
Q}\left[(1-\alpha)N^{1/3}\right]$} (Fig.~3).

The threshold window is larger than the canonical $N^{-1/2}$ estimate
arising either from the standard large-population analysis
\cite{ngv,dss} or from the widely-used deterministic SIR ordinary
differential equations \cite{jdm}, describing the evolution of the
average susceptible and infected populations \cite{bk}. Moreover, the
related SI (sometimes also termed SIS) model, where a recovered
individual immediately becomes susceptible, is characterized by the
simpler behavior $n_*\sim N$ and $\langle n\rangle \sim N^{1/2}$;
finite size effects are not as pronounced because there is no
depletion of the susceptible reservoir.

The scaling laws for the outbreak size have direct implications
concerning the dynamics and in particular, the duration of infection
processes near the epidemic threshold. To obtain these scaling laws,
we again consider first the infinite population limit. At the epidemic
threshold, $\alpha=1$, infection and recovery occur with equal
probabilities and therefore, the average number of infected
individuals is conserved, $I(t)=1$. The probability $P_i(t)$ that
there are $i$ infected individuals at time $t$ satisfies
\begin{equation}
\frac{d}{dt}P_i=(i+1)P_{i+1}+(i-1)P_{i-1}-2iP_i
\end{equation}
together with the initial condition $P_i(0)=\delta_{i,1}$. The
distribution is geometric, $P_i(t)=t^{i-1}(1+t)^{-(i+1)}$ \cite{bk,edg}
for $i\geq 1$, and $P_0(t)=t(1+t)^{-1}$ for $i=0$. Therefore, the
survival probability of the outbreak, i.e., the probability that the
outbreak is still active at time $t$ is simply
\begin{equation}
\label{Pt} 
P(t)=(1+t)^{-1}
\end{equation}
since $P(t)=1-P_0(t)$.  Restricting attention to active outbreaks, the
average number of infected individuals grows linearly with time
$\langle i\rangle=I(t)/P(t)=1+t$. Consequently, the typical number of
recovered individuals $r\sim \int_0^t dt'\,(1+t')$ grows quadratically
with time: $r\sim t^2$.

\begin{figure}[t]
  \centerline{\epsfxsize=8cm\epsfbox{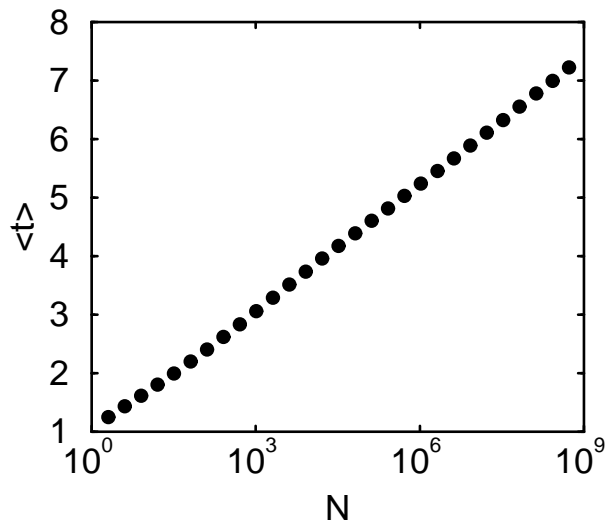}}
 \caption{The average outbreak duration at the epidemic threshold
 versus the population size. Simulation results, obtained from an
 average over $10^9$ realizations are consistent with the theoretical
 prediction (\ref{duration-ave}). A best fit to $\langle t\rangle =
 \beta\ln N$ yields $\beta=0.32\pm 0.01$.}
 \label{tav}
\end{figure}

For finite populations, the probability that the outbreak is still
alive at time $t$ decays as $P(t,N)\sim t^{-1}$ up to the maximal time
scale $t\ll t_*$.  The survival probability is sharply suppressed for
times larger than the maximal time.  The maximal duration of outbreaks
is estimated by equating the time dependent outbreak size $n\sim r
\sim t^2$ with the maximal outbreak size $n_*\sim N^{2/3}$. Therefore,
\begin{equation}
\label{duration-max}
t_*\sim N^{1/3}.
\end{equation}

The maximal duration of outbreaks greatly exceeds both the typical
duration that is of the order of one and the average duration of an
outbreak $\langle t\rangle$ which exhibits an interesting logarithmic
growth. To derive the logarithmic law, we first note that, by
definition, the average duration of an outbreak is $\langle
t\rangle=\int_0^{\infty}\, dt\, t\, \frac{d}{dt}P(t,N)$. Using the
infinite population result (\ref{Pt}) and integrating up to $t_*$ that
plays the role of a cutoff, we get
\begin{equation}
\label{duration-ave}
\langle t\rangle\simeq \frac{1}{3}\ln N.
\end{equation}
Numerical simulations confirm this behavior (Fig.~4). The probability
distribution for the duration of outbreaks also follows a
population-size independent law: \hbox{$P(t,N)/P(t)\to {\cal
P}\left(t/N^{1/3}\right)$} as shown in Fig.~5. However, the
convergence to this law is not uniform: it is slow for short durations
but fast at large durations.

\begin{figure}[t]
\centerline{\epsfxsize=8cm\epsfbox{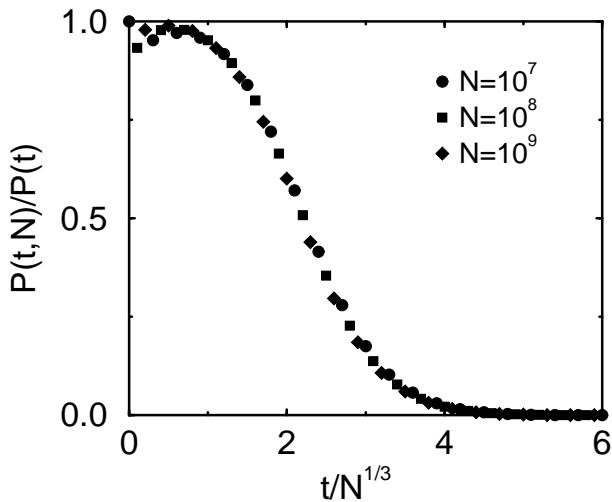}} 
\caption{The survival probability at the epidemic threshold. Shown is
the normalized survival probability $P(t,N)/P(t)$ versus the
normalized duration time $t/N^{1/3}$. The data corresponds to an
average over $10^8$ realizations.}
\label{st}
\end{figure}

In summary, we found that outbreaks in the vicinity of the epidemic
threshold have a distinct size, characterized by a distinct power-law
dependence of the population size.  This behavior describes a range of
infection processes in the vicinity of the epidemic threshold. The
size of this threshold window is larger than expected from the
traditional large system size analysis techniques or from the
deterministic description.  We conclude that statistical fluctuations
and finite population effects are most pronounced and may be quite
subtle near the epidemic threshold.

The scaling laws have concrete implications regarding the
computational complexity of near-threshold infection
processes. Typically, one has to compute $P_{i,r}$, the probability
that there are $i$ infected individuals and $r$ removed individuals
from the master equations.  Although there are $N^2$ such coupled
ordinary differential equations, the scaling laws $i\sim N^{1/3}$ and
$r \sim N^{2/3}$ imply that the number of relevant equations is much
smaller and scales only linearly with the population size.

Several questions arise, e.g., what is the shape of the scaling
functions ${\cal G}\left(n/N^{2/3}\right)$ and ${\cal
P}\left(t/N^{1/3}\right)$ characterizing the size and duration of
outbreaks near the epidemic threshold? Analytical determination of
these functions is very challenging as it requires treatment of the
full master equations describing the stochastic infection process
\cite{ntjb}, that is, the distribution $P_{i,r}(t,N)$ is needed
\cite{bk}.

Further related problems include the corresponding near-threshold
scaling laws for spatial epidemic models, where the geometry and the
spatial structure of the infected domain play a role
\cite{sa,pg,ad1,wss}, and infection processes on networks
\cite{pv,mejn}.

\acknowledgments We are grateful to Aric Hagberg for initial
collaboration on this work. We also thank Gary Doolen, Hans
Frauenfelder, and Sergei Rudchenko for careful reading of the manuscript.
This research was supported in part by DOE(W-7405-ENG-36).

\end{document}